\documentstyle[12pt]{article}

\catcode`\@=11
\long\def\@makefntext#1{
\protect\noindent \hbox to 3.2pt {\hskip-.9pt
$^{{\ninerm\@thefnmark}}$\hfil}#1\hfill}                

\def\@makefnmark{\hbox to 0pt{$^{\@thefnmark}$\hss}}  
	
\def\ps@myheadings{\let\@mkboth\@gobbletwo
\def\@oddhead{\hbox{}
\rightmark\hfil\ninerm\thepage}
\def\@oddfoot{}\def\@evenhead{\ninerm\thepage\hfil
\leftmark\hbox{}}\def\@evenfoot{}
\def\sectionmark##1{}\def\subsectionmark##1{}}

\renewcommand{\thefootnote}{\fnsymbol{footnote}}

\newcounter{sectionc}\newcounter{subsectionc}\newcounter{subsubsectionc}
\renewcommand{\section}[1] {\vspace*{0.6cm}\addtocounter{sectionc}{1}
\setcounter{subsectionc}{0}\setcounter{subsubsectionc}{0}\noindent
	{\normalsize\bf\thesectionc. #1}\par\vspace*{0.4cm}}
\renewcommand{\subsection}[1] {\vspace*{0.6cm}\addtocounter{subsectionc}{1}
	\setcounter{subsubsectionc}{0}\noindent
	{\normalsize\it\thesectionc.\thesubsectionc. #1}\par\vspace*{0.4cm}}
\renewcommand{\subsubsection}[1]
{\vspace*{0.6cm}\addtocounter{subsubsectionc}{1}
	\noindent {\normalsize\rm\thesectionc.\thesubsectionc.\thesubsubsectionc.
	#1}\par\vspace*{0.4cm}}

\newcounter{appendixc}
\newcounter{subappendixc}[appendixc]
\newcounter{subsubappendixc}[subappendixc]

\renewcommand{\appendix}[1] {\vspace*{0.6cm}
	\refstepcounter{appendixc}
	\setcounter{figure}{0}
	\setcounter{table}{0}
	\setcounter{equation}{0}
	\renewcommand{\thefigure}{\Alph{appendixc}.\arabic{figure}}
	\renewcommand{\thetable}{\Alph{appendixc}.\arabic{table}}
	\renewcommand{\theappendixc}{\Alph{appendixc}}
	\renewcommand{\theequation}{\Alph{appendixc}.\arabic{equation}}
	\noindent{\bf Appendix \theappendixc #1}\par\vspace*{0.4cm}}

\def\abstracts#1{{
	\centering{\begin{minipage}{12.2truecm}\footnotesize\baselineskip=12pt\noindent
	\centerline{\footnotesize ABSTRACT}\vspace*{0.3cm}
	\parindent=0pt #1
	\end{minipage}}\par}}

\renewenvironment{thebibliography}[1]
	{\begin{list}{\arabic{enumi}.}
	{\usecounter{enumi}\setlength{\parsep}{0pt}
\setlength{\leftmargin 1.25cm}{\rightmargin 0pt}
	 \setlength{\itemsep}{0pt} \settowidth
	{\labelwidth}{#1.}\sloppy}}{\end{list}}

\newcounter{itemlistc}
\newcounter{romanlistc}
\newcounter{alphlistc}
\newcounter{arabiclistc}

\newcommand{\fcaption}[1]{
	\refstepcounter{figure}
	\setbox\@tempboxa = \hbox{\footnotesize Fig.~\thefigure. #1}
	\ifdim \wd\@tempboxa > 6in
	   {\begin{center}
	\parbox{6in}{\footnotesize\baselineskip=12pt Fig.~\thefigure. #1}
	    \end{center}}
	\else
	     {\begin{center}
	     {\footnotesize Fig.~\thefigure. #1}
	      \end{center}}
	\fi}

\newcommand{\tcaption}[1]{
	\refstepcounter{table}
	\setbox\@tempboxa = \hbox{\footnotesize Table~\thetable. #1}
	\ifdim \wd\@tempboxa > 6in
	   {\begin{center}
	\parbox{6in}{\footnotesize\baselineskip=12pt Table~\thetable. #1}
	    \end{center}}
	\else
	     {\begin{center}
	     {\footnotesize Table~\thetable. #1}
	      \end{center}}
	\fi}

\def\@citex[#1]#2{\if@filesw\immediate\write\@auxout
	{\string\citation{#2}}\fi
\def\@citea{}\@cite{\@for\@citeb:=#2\do
	{\@citea\def\@citea{,}\@ifundefined
	{b@\@citeb}{{\bf ?}\@warning
	{Citation `\@citeb' on page \thepage \space undefined}}
	{\csname b@\@citeb\endcsname}}}{#1}}

\newif\if@cghi
\def\cite{\@cghitrue\@ifnextchar [{\@tempswatrue
	\@citex}{\@tempswafalse\@citex[]}}
\def\citelow{\@cghifalse\@ifnextchar [{\@tempswatrue
	\@citex}{\@tempswafalse\@citex[]}}
\def\@cite#1#2{{$\null^{#1}$\if@tempswa\typeout
	{IJCGA warning: optional citation argument
	ignored: `#2'} \fi}}

 1
 1
 1

\font\ninerm=cmr9

\begin{document}
\hfill KL--TH--96/1 \\[5mm]
\centerline{\normalsize\bf INSTANTON INDUCED TUNNELING AMPLITUDE AT EXCITED STATES}\\
\centerline{\normalsize{\bf WITH THE LSZ METHOD}}
\baselineskip=16pt

\vspace*{0.6cm}
\centerline{\footnotesize J.-G. ZHOU,J.-Q.LIANG,J.BURZLAFF\footnote{On leave from Department
of Mathematics, Dublin City University, Dublin, Ireland} and H.J.W.M\"ULLER-KIRSTEN}
\baselineskip=13pt
\centerline{\footnotesize\it Department of Physics, University of Kaiserslautern, P.O.Box 3049}
\baselineskip=12pt
\centerline{\footnotesize\it 67653 Kaiserslautern, Germany}
\centerline{\footnotesize E-mail: jgzhou@gypsy.physik.uni-kl.de}
\vspace*{0.3cm}

\vspace*{0.9cm}
\abstracts{Quantum tunneling between degenerate ground states through
the central barrier of a potential is extended to excited states with the
instanton method.  This extension is achieved with the help of
an LSZ reduction technique as in field theory and may be of
importance in the study of macroscopic quantum phenomena in
magnetic systems.}
\vspace*{0.9cm}
\baselineskip=16pt

\indent
Quantum effects on the macroscopic scale have attracted considerable attention
in recent years owing mainly to the development of technology in
mesoscopic physics. Legget et al.\cite{1} predicted that the most 
intriguing quantum effect which could take place
on the macroscopic scale is quantum tunneling.  Macroscopic
magnetisation tunneling is a subject which is being investigated
extensively and is of growing interest.  There are two types of
macroscopic quantum phenomena in magnetic systems. One appears
when either a giant spin or (in bulk material) a domain wall is
tunneling between degenerate states, which is the situation
of macroscopic quantum coherence\cite{1,2,3}. Quantum tunneling
in this case is dominated by the instanton configuration with 
nonzero topological charge, and the tunneling results in the level splitting.
The second phenomenon appears in macroscopic quantum
tunneling dominated by the socalled bounce configuration\cite{4}
with zero topological charge which leads to the decay of metastable states\cite{5}.
In the case of the quantum depinning of a domain wall, pinned by defects,
the position of the wall at the pinning centre becomes metastable
in the external magnetic field, and the position tunnels out of the
local minimum\cite{6,7}.  However, the usual instanton method only provides
the amplitude for tunneling between ground states.  It is therefore
of interest to evaluate also the tunneling effect between excited
states.  Motivated by the study
of baryon- and lepton-number violation  at high energy, recently\cite{8}
a periodic (or nonvacuum) instanton configuration has been found and 
used to evaluate the quantum tunneling at high energy.  In this note 
we present an approach with the LSZ reduction technique in field theory
to calculate the tunneling amplitude between asymptotically degenerate 
excited states. In the low energy limit the result
is equally good as that given in the literature\cite{8} by use of the
periodic instanton.  However, the procedure employed here avoids
the divergence difficulty of the Feynman propagator between two turning 
points involved in the periodic instanton method\cite{8}, and is\
more adequate for the low temperature case.

 The idea of a tunneling 
transition from one side of a
potential barrier to the other has recently also been linked with the LSZ reduction mechanism of
a transition from asymptotic in-states to asymptotic out-states\cite{9,10,11}. But although this
idea is very attractive and worth pursuing it has not been studied in detail. 
In the following we therefore go beyond a cursory employment of the method and
use the LSZ reduction procedure in a modified way in order to calculate the tunneling amplitude in the one-instanton
sector for the sine-Gordon potential including the contribution of quantum fluctuations 
up to the one-loop approximation.  This example is of particular interest in the
context of macroscopic tunneling phenomena in magnetic systems.

\setcounter{footnote}{0}
\renewcommand{\thefootnote}{\alph{footnote}}

We recall first the  case of a one-dimensional harmonic oscillator described by the Hamiltonian
\begin{equation}
H = \frac{1}{2}p^2 + \frac{1}{2}\omega^2 q^2
\label{1}
\end{equation}
for mass $m = 1 $ and $ \hbar = 1$. Here $q$ and $p$ are dynamical observables which become
operators when subjected to the Heisenberg algebra of ordinary canonical quantisation.  The
solution of the Heisenberg equation of motion,
$ \ddot{q} +\omega^2 q = 0$, then becomes
\begin{equation}
q(t) = \frac{1}{\sqrt{2\omega}}[a e^{-i\omega t} + a^{\dagger} e^{i\omega t}]
\label{2}
\end{equation}
where $a, a^{\dagger}$ are time-independent operators defined by the initial ($t = 0$) values of
$q$ and $p$, i.e.
\begin{equation}
q(0) = \frac{1}{\sqrt{2\omega}}[a + a^{\dagger}],  \\  p(0)= \frac{-i\omega}{\sqrt{2\omega}}[a - a^{\dagger}]
\label{3}
\end{equation}
In fact, a rotation of $q(t), \frac{p(t)}{\omega}$ through angle $\omega t$ transforms these back
to their initial values.  The operators $a, a^{\dagger} $ can be obtained from $q(t), p(t) = \dot{q}(t)$.
Thus
\begin{eqnarray}
a^{\dagger} &=& -\frac{i}{\sqrt{2\omega}}e^{-i\omega t}[\dot{q}(t) + i\omega q(t)]\nonumber\\
&\equiv & -\frac{i}{\sqrt{2\omega}} e^{-i\omega t} \stackrel{\leftrightarrow}{\frac{\partial}{\partial t}}q(t)\nonumber\\
\label{4}
\end{eqnarray}
and $a$ follows with complex conjugation.  One should note the extra minus sign in the
definition of the symbol $\stackrel{\leftrightarrow}{\frac{\partial}{\partial t}}$ when acting to the left.  
Operators of this type are well-known in the literature\cite{12}.

We now consider the (1+0)-dimensional theory defined by the Lagrangian
\begin{equation}
{\cal L} = \frac{1}{2} (\frac{d\phi}{dt})^2 - V(\phi)
\label{5}
\end{equation}
with the sine-Gordon potential
\begin{equation}
V(\phi) = \frac{1}{g^2} (1 + \cos g\phi)
\label{6}
\end{equation}
It is well-known that the classical equation of motion of this theory, for energy $E = 0$,
with Euclidean time $\tau = i t$, 
i.e.
\begin{equation}
\frac{1}{2}(\frac{d\phi}{d\tau})^2 - V(\phi) = 0
\label{7}
\end{equation}
possesses the instanton solution
\begin{equation}
\phi_c = \frac{2}{g} \sin^{-1}[\tanh (\tau-\tau_0)]
\label{8}
\end{equation}
as a nontrivial (i.e. $\tau$-dependent) and topological vacuum.  Expanding $V(\phi)$ around
its principal minima at $\phi = \pm \frac{\pi}{g}$ we obtain
\begin{equation}
V(\phi) = \frac{1}{2}[\phi - (\pm \frac{\pi}{g})]^2 + ...
\end{equation}
Comparison with the oscillator case discussed above shows that the oscillator frequency
$\omega$ around a minimum is $\omega = 1$. Crucial aspects of the LSZ procedure \cite{13}
are its asymptotic conditions which require the theory to have an interpretation in terms of
observables for stationary incoming and outgoing states. We can simulate such a situation here
artificially by imagining the central barrier of the potential to be extremely high 
and the neighbouring wells ``--'' and ``+'' on either side as extremely far apart.  We
therefore construct appropriate functions $\phi_{\pm}(\tau)$ which become
oscillator-like in the limits $\tau \rightarrow \pm\infty$ respectively.   These functions must be
formulated in terms of the instanton $\phi_c$ which provides the interaction, i.e. 
interpolation between the asymptotic states.  We define therefore the real ``interaction fields''
\begin{equation}
\phi_{\mp}:= \frac{\pi}{g} \pm \phi_c
\label{9}
\end{equation}
which are such that
\begin{equation}
\lim_{\tau\rightarrow \pm\infty} \phi_{\pm} = 0
\end{equation} and define
\begin{eqnarray}
a_{\pm}^{\dagger}&:=&\sqrt{2}e^{-\tau}\stackrel{\leftrightarrow}{\frac{\partial}{\partial\tau}}\phi_{\pm}(\tau),\\
a_{\pm}&:=&-\sqrt{2}e^{\tau}\stackrel{\leftrightarrow}{\frac{\partial}{\partial\tau}}\phi_{\pm}(\tau)\ 
\end{eqnarray}
These ``creation'' and ``annihilation fields'' of effective or quasi--bosons in wells
``$+$'' and ``$-$'' are related to the ``interaction fields'' through the formula
\begin{equation}
2\phi_{\pm}=\frac{1}{\sqrt{2}}[a_{\pm}e^{-\tau} + a_{\pm}^{\dagger} e^{\tau}]
\label{11}
\end{equation}
Since differentiation of the defining expressions of $\phi_{\pm}$ in eq.(10) gives
\begin{equation}
\frac{\partial\phi_{\pm}}{\partial\tau} = {\mp}\frac{2}{g\cosh (\tau-\tau_{0})} \stackrel{\tau \rightarrow\pm\infty}    
{\longrightarrow}
\mp\frac{4}{g} e^{\mp\tau}
\end{equation}
we see that
\begin{eqnarray}
a_{\pm}^{\dagger} \stackrel{\tau \rightarrow{-\infty}}{\longrightarrow}\mp \frac{4\sqrt 2}{g}\\
a_{\pm}\stackrel{\tau \rightarrow{+\infty}}{\longrightarrow}\pm \frac{4\sqrt 2}{g}
\end{eqnarray}
and
\begin{equation}
\lim_{\tau \rightarrow -\infty} a_{\pm}(\tau) = 0, \lim_{\tau \rightarrow +\infty} a_{\pm}^{\dagger}
(\tau) = 0
\end{equation}
i.e. comparing $q$ of the oscillator case with $\phi_{\pm}$ we see that $a_{oscillator} = \frac{4}
{g\sqrt 2}$. Just as quantisation of the oscillator case is achieved by raising $a^{\dagger}(t),a(t),q(t),...$
to operators $\hat{a}^{\dagger}(t),\hat{a}(t),\hat{q}(t),...$, so here quantisation implies raising $a^{\dagger}(\tau),
a(\tau),\phi (\tau),..$ to operators $\hat{a}^{\dagger}(\tau),\hat{a}(\tau), \hat{\phi} (\tau),...$.   
If $|0>$ is the perturbation theory vacuum state in either well of the potential (i.e. when the
central barrier is infinitely high), the state $ a_-^\dagger (\tau)
|0> $ with $\tau \rightarrow -\infty$ represents a one-quasi-boson oscillator state in that well.
We consider this as the one-quasi-boson in-state and hence write the amplitude for the
one boson transition from one side (``--") of the central barrier to the other (``+")
\begin{displaymath}
A_{f,i} = <+,1|1,-> = S_{f,i}e^{-2T}
 \end{displaymath}
with the S--matrix element $S_{f,i}$ is given by
\begin{equation}
S_{f,i}=\lim_{\tau \rightarrow -\infty, \tau'\rightarrow +\infty}
	    <0| \hat{a}_+(\tau')\hat{ a}_-^\dagger(\tau)|0>\\ 
\label{13}
\end{equation}
where $\lim_{\tau \rightarrow -\infty}\hat{a}_{\pm}(\tau)|0> = 0 $. Inserting (12), (13) we have
\begin{equation}
S_{f,i}= \lim_{\tau \rightarrow -\infty,\tau' \rightarrow +\infty} 
(-\sqrt{2} e^{\tau'}\stackrel{\leftrightarrow}{\frac{\partial}{\partial\tau'}})
(\sqrt{2}e^{-\tau} \stackrel{\leftrightarrow}{\frac{\partial}{\partial\tau}})
G(\tau',\tau)
\label{14}
\end{equation}
where $ G $ is the Green's function
\begin{equation}
G(\tau',\tau) = <0|\hat{\phi_+}(\tau') \hat{\phi_-}(\tau)|0>
\label{15}
\end{equation}
The differentiations in $S_{f,i}$ imply
\begin{equation}
S_{f,i}=\lim_{\tau \rightarrow -\infty,\tau' \rightarrow +\infty}-{2e^{- \tau}e^{\tau'}}
[(\frac{\partial^2 G}{\partial \tau  \partial \tau'} +\frac{\partial G}{\partial \tau'})
-(\frac {\partial G}{\partial \tau} + G)]
\label{16}
\end{equation}
We evaluate G by inserting complete sets of states of final and initial field configurations
$\phi_f, \phi_i $. Thus
\begin{eqnarray}
G(\tau',\tau) &= &<0|\hat{\phi_+}(\tau') \hat{\phi_-}(\tau)|0>\nonumber\\
&=&\int d\phi_f d\phi_i <0|\phi_f><\phi_f|\hat{\phi_+}(\tau')\hat{\phi_-}(\tau)|\phi_i><\phi_i|0>\nonumber\\
&=&\int d\phi_f d\phi_i <0|\phi_f><\phi_i|0>\phi_+(\tau')\phi_-(\tau) <\phi_f|\phi_i>
\label{17}
\end{eqnarray}
with
\begin{equation}
\hat{\phi}_{-}(\tau)|\phi_{i}> = \phi_{-}(\tau)|\phi_{i}>
\label{17.1}
\end{equation}
Here $<0|\phi_f>, <\phi_i|0>$ are degenerate ground state wave functions in the two wells
which we write
\begin{equation}
<0|\phi_f> \equiv \psi_0(\phi_f),\;\;  <\phi_i|0> \equiv \psi_0(\phi_i) 
\end{equation}
and
$<\phi_f|\phi_i> \equiv <\phi_f,\tau'|\phi_i,\tau> $ is the propagator
\begin{equation}
K(\phi_f,\tau';\phi_i,\tau) \equiv \int_{\phi_i}^{\phi_f} {\cal D}{\phi}e^{-S} 
\label{18}
\end{equation}
Considering the limits $\tau \rightarrow -\infty,\tau' \rightarrow \infty$, we can write
\begin{equation}
K:=\lim_{\tau \rightarrow -\infty,\tau'\rightarrow \infty,\phi_i \rightarrow -\frac{2}{g},\phi_f
\rightarrow \frac{2}{g}} K(\phi_f,\tau';\phi_i,\tau)
\label{18.1}
\end{equation}
Then $K$ representing the propagator can be removed from the integral and
\begin{eqnarray}
\lim_{\tau \rightarrow -\infty,\tau' \rightarrow \infty} G(\tau', \tau) &=&\lim_{\tau\rightarrow -\infty,
\tau'\rightarrow \infty} K \phi_+(\tau')
\phi_-(-\tau)\int d\phi_f d\phi_i \psi_0(\phi_f) \psi_0(\phi_i)\\
& \approx &\lim_{\tau\rightarrow -\infty,\tau'\rightarrow \infty} 2\sqrt{\pi} K\phi_+(\tau')\phi_-(-\tau)
\label{18.2}
\end{eqnarray}
\noindent
Since
\begin{equation}
\lim_{\tau\rightarrow - \infty}\phi_-(\tau) = 0,\;\; \lim_{\tau'\rightarrow +\infty} \phi_+(\tau') = 0
\end{equation}
we see that
\begin{equation}
\lim_{\tau' \rightarrow +\infty, \tau \rightarrow - \infty} G(\tau',\tau) = 0
\label{19}
\end{equation} 
In the combined limits $\tau \rightarrow - \infty, \tau' \rightarrow +\infty$ also
$\frac {\partial G}{\partial \tau} $ and $\frac {\partial G}{\partial \tau'} \rightarrow 0$. 
Thus in (22) the only nonvanishing contribution in these limits results from the second
derivative.  Then
\begin{eqnarray}
(\frac {\partial^2 G}{\partial \tau \partial\tau'})_{\stackrel{\tau \rightarrow -\infty}{\tau' \rightarrow
+\infty}}
&=& 2\sqrt {\pi} K(\frac {\partial \phi_+(\tau')}{\partial \tau'})_{\tau'\rightarrow \infty}
(\frac {\partial \phi_-(\tau)}{\partial \tau})_{\tau \rightarrow -\infty}
\nonumber\\
&=& -2\sqrt {\pi} K (\frac {\partial \phi_c(\tau')}{\partial \tau'})_{\tau' \rightarrow \infty}
(\frac {\partial \phi_c(\tau)}{\partial\tau})_{\tau \rightarrow -\infty} \nonumber\\
&=& -2\sqrt{\pi} K (\frac{4}{g})^2 (e^{-\tau'}e^{\tau})_{\stackrel{\tau' \rightarrow \infty}{\tau \rightarrow
-\infty}}
\label{20}
\end{eqnarray}
Inserting the result into $S_{f,i}$ and hence into $A_{f,i},$ and taking the limits
with $2T=\tau'-\tau \rightarrow \infty$ 
we obtain 
\begin{equation}
A_{f,i} = 2\sqrt{\pi}. K .(\frac{4}{g})^2e^{-2T}
\label{21}
\end{equation}
The factor $(\frac{4}{g})^2$ is the product
of asymptotic 
amplitudes of the two operators as can be seen from (15).  
The next step is to calculate the propagator $K$ with the instanton method. To this end we expand
$\phi(\tau)$ around the instanton trajectory $\phi_c(\tau)$ such that
\begin{equation}
\phi(\tau) = \phi_c + \chi (\tau)
\label{21.1}
\end{equation}
where $\chi (\tau) $ denotes the quantum fluctuation.  The propagator is now evaluated in
the one-loop approximation, and the result is found to be
\begin{equation}
K = \frac{2T}{\pi} e^{-\frac{8}{g^2}}.e^{-T}.\frac{4}{g}
\label{21.2}
\end{equation}
The tunneling amplitude then reads
\begin{equation}
A_{f,i}= \frac{2Te^{-\frac{8}{g^2}}}{\sqrt{\pi} g} 4 e^{-3T} (\frac{4\sqrt{2}}{g})^2 \equiv A_{f,i}^{1\rightarrow 1}
\label{22}
\end{equation}
where $2T = \tau' -\tau$.  In this result the factor $2T$ can be interpreted heuristically
as representing energy conservation in the sense of the relation
\begin{equation}
2\pi \delta (E_ f -E_i) \stackrel{T\rightarrow \infty}{=}\int_{-T}^{T} dt e^{i(E_f-E_i)t}
\end{equation}
or
\begin{equation}
2\pi \delta (0) \stackrel{T\rightarrow \infty}{=}\int_{-T}^{T} dt = 2T
\label{23}
\end{equation}
The factor $e^{-3T} = [e^{-(n+\frac{1}{2})2T}]_{n=1}$(the one-quasi-boson case)
 represents in Minkowski
time the free-field evolution.  
We can obtain the corresponding amplitude for the transition
between the nth excited states on either side of the barrier in the one-instanton
sector by evaluating
\begin{equation}
A_{f,i}^{n\rightarrow n}=\frac{1}{n!}\prod_{i=1}^{n} \lim_{\tau_i\rightarrow -\infty,
\tau_i'\rightarrow \infty}(-\sqrt{2}e^{\tau_i'}\stackrel{\leftrightarrow}
{\frac{\partial}{\partial\tau_i'}})(\sqrt{2}e^{-\tau_i}
\stackrel{\leftrightarrow}{\frac{\partial}{\partial \tau_i}}) 
G(\tau_1',\tau_2',...,\tau_n';\tau_1,\tau_2,...,\tau_n)\nonumber\\
\label{25}
\end{equation}
and it may be surmised that with calculations analogous to those above
\begin{equation}
A_{f,i}^{n \rightarrow n}= 2T \frac{e^{-\frac{8}{g^2}}}{\sqrt{\pi} g}\frac{4 e^{-2E_{cl}T}}{n!}
(\frac{4\sqrt{2}}{g})^{2n}
\label{26}
\end{equation}
where now $E_{cl} = (n+\frac{1}{2})$.
Without tunneling the eigenstates in neighbouring wells are degenerate.  The
degeneracy is removed by the small tunneling effect which leads to the
level splitting while the levels extend to bands due to the translational
symmetry of the sine-Gordon potential \cite{8}.  The relation between the tunneling
amplitude and the level splitting is given by \cite{8} 
\begin{equation}
A^{n \rightarrow n} = e^{-2E_{cl}T} \sinh (2\triangle \epsilon_{n}T)
\label{27}
\end{equation}
where $\triangle \epsilon_n$ denotes the level splitting of the $n$-th energy
eigenvalue.  The one--instanton sector of the tunneling
amplitude, eq.(37), does not show the proper hyperbolic function which, however, 
can be obtained by taking into account the contributions of the
infinite number of instanton--antiinstanton pairs\cite{8}.  The
tunneling effect is, of course, extremely small.  One may expand the 
hyperbolic function in eq. (42) in rising powers of $2\triangle \epsilon_n$.  
Up to the first order we have
\begin{equation}
\triangle \epsilon_{n} \approx \frac {4e^{-\frac {8}{g^2}}}{\sqrt{\pi} g n!}(\frac {4\sqrt{2}}{g})^{2n}
\label{28}
\end{equation}
which reduces to the correct value of the level splitting of the ground state when $n = 0 $
\cite{8}.
To our knowledge the level splitting due to tunneling at excited states within the
framework of the instanton method applied to the sine-Gordon potential has not
been reported previously in the literature, and the explicit calculation of the one-loop
correction in this case is also not wellknown, and of interest in the study of macroscopic 
quantum tunneling in magnetic systems.

The calculation of the contributions of instanton--antiinstanton pairs 
 and the application to macroscopic tunneling in a specific system will
will be considered elsewhere.
\newline\\

\noindent {\bf Acknowledgements} \\[1mm]
J.-G.Z. is indebted to the A.v.Humboldt Foundation for the award of a Fellowship,and
J.-Q.L. to the Deutsche Forschungsgemeinschaft 
 and J.B.
to the sponsors of Irish-German Scientific Cooperation (GKSS,Geesthacht, Germany and
EOLAS,Dublin,Ireland) for supporting their visits to Kaiserslautern.
\\[25mm]

\end{document}